\documentclass[12pt, a4paper]{article}

\usepackage[height=22cm,width=16cm, centering]{geometry}
\usepackage{setspace}

\usepackage[utf8]{inputenc}

\usepackage{amsmath, comment}
\usepackage{amssymb}
\usepackage{graphicx}
\usepackage{color}
\usepackage[sort&compress, numbers, merge]{natbib}

\definecolor{purple}{rgb}{0.5 ,0, 0.7}
\definecolor{bluegreen}{rgb}{0, 0.45, 0.35}
\definecolor{sakura}{rgb}{1 ,0.52, 0.74}
\definecolor{wakakusa}{rgb}{0.45 ,0.74, 0}
\definecolor{brown}{rgb}{0.48 ,0.23, 0}
\definecolor{skyblue}{rgb}{0.21 ,0.7, 1.}
\definecolor{purplegray}{rgb}{0.35,0.35,0.73}
\usepackage{hyperref}
\hypersetup{colorlinks=true, linkcolor=bluegreen, citecolor=purple, urlcolor=blue}

\makeatletter
\let\MYcaption\@makecaption
\makeatother

\usepackage{subcaption}
\makeatletter
\let\@makecaption\MYcaption
\makeatother

\allowdisplaybreaks

\begin{document}

\begin{titlepage}
\begin{center}
\leavevmode \\
\vspace{ 0cm}

 \hfill{\small KEK-TH-2260}\\
 \hfill{\small KEK-Cosmo-0263}\\
\hfill {\small CTPU-PTC-20-22}\\

\noindent
\vskip 1.5 cm

  {\LARGE Solar-Mass Primordial Black Holes Explain  \\  \vskip 1.8 mm  NANOGrav Hint of Gravitational Waves}
  
\vglue .6in

{
Kazunori Kohri$^{a, b, c}$ and Takahiro Terada$^{d}$
}

\vglue.3in

\small
\textit{ 
${}^{a}$ \,  Institute of Particle and Nuclear Studies, KEK,
1-1 Oho, Tsukuba, Ibaraki 305-0801, Japan \\
${}^{b}$ \,The Graduate University for Advanced Studies (SOKENDAI), \\
1-1 Oho, Tsukuba, Ibaraki 305-0801, Japan\\
${}^{c}$ \,Kavli Institute for the Physics and Mathematics of the Universe (WPI),\\
University of Tokyo, Kashiwa 277-8583, Japan \\
${}^{d}$ \,Center for Theoretical Physics of the Universe,\\ Institute for Basic Science (IBS),
  Daejeon, 34126, Korea 
}
\normalsize

\end{center}

\vglue 0.6in

\begin{abstract}
The NANOGrav collaboration for the pulsar timing array (PTA) observation recently announced evidence of an isotropic stochastic process, which may be the first detection of the stochastic gravitational-wave (GW) background.  We discuss the possibility that the signal is caused by the second-order GWs associated with the formation of solar-mass primordial black holes (PBHs).  This possibility can be tested by future interferometer-type GW observations targeting the stochastic GWs from  merger events of solar-mass PBHs as well as by updates of PTA observations.  
\end{abstract}
\end{titlepage}

\newpage

\section{Introduction} \label{sec:intro}

Gravitational-wave (GW) astronomy started with the successful observations of GWs from merger events of binary black holes by LIGO/Virgo collaborations~\cite{LIGOScientific:2018mvr}.  GWs are also a valuable probe for the early Universe cosmology and particle physics.  In particular, interests in primordial black holes (PBHs)~\cite{Hawking:1971ei, Carr:1974nx, Carr:1975qj} were reactivated after the first detection of GWs~\cite{Bird:2016dcv, Clesse:2016vqa, Sasaki:2016jop}.  In the PBH scenario, GWs can be emitted not only from the merger of binary PBHs but also from the enhanced curvature perturbations that form PBHs~\cite{Saito:2008jc, Saito:2009jt, Bugaev:2009zh}.  This is due to the scalar-tensor mode couplings appearing at the second-order of the cosmological perturbation theory~\cite{tomita1967non, Matarrese:1993zf, Matarrese:1997ay, Ananda:2006af, Baumann:2007zm, Assadullahi:2009nf}.  It is interesting that we can indirectly probe physics of inflation by probing the primordial scalar (curvature/density) perturbations inferred from the second-order GWs and PBH abundances~\cite{Alabidi:2012ex, Alabidi:2013lya, Orlofsky:2016vbd, Inomata:2018epa, Byrnes:2018txb, Ben-Dayan:2019gll, Unal:2020mts}.

Recently, the North American Nanohertz Observatory for Gravitational Waves (NANOGrav) released its 12.5-year pulsar timing array (PTA) data~\cite{Arzoumanian:2020vkk}.  They search for an isotropic stochastic GW background by analyzing the cross-power spectrum of pulsar timing residuals.  They reported evidence of a stochastic process, parametrized as a power-law, whose amplitude and slope are common among pulsars.  The significance of the quadrupole nature in the overlap reduction function is not conclusive, whereas the monopole and dipole are relatively disfavored.   This implies that the NANOGrav collaboration might have detected an astrophysical or cosmological stochastic GW background.  

It should be noted that the NANOGrav 12.5-yr signal strength is greater than the upper bound derived in their previous 11-yr result~\cite{Arzoumanian:2018saf} as well as that in Parkes PTA (PPTA)~\cite{Shannon:2015ect} (see Ref.~\cite{Chen:2019xse} for the NANOGrav 11-yr constraints on PBHs and also Ref.~\cite{Cai:2019elf} related particularly to European PTA (EPTA) constraints~\cite{Lentati:2015qwp}).    
This apparent tension is explained primarily by the different choices of the Bayesian priors~\cite{Arzoumanian:2020vkk, Hazboun:2020kzd}, so all analyses can be correct given their assumptions including the priors.  
Specifically, the most relevant prior is on the amplitude of the red noise component associated with each pulsar.
Previous PTA analyses used the uniform prior in the linear scale, whereas the NANOGrav 12.5-yr analyses used the uniform prior in the log scale.
The effect of the difference is studied in detail in Ref.~\cite{Hazboun:2020kzd}, and they found that the injected GW signal in their simulations tends to be absorbed by the red noise component more easily in the case of the (linearly) uniform prior.  Moreover, the 95\% confidence-level upper bound on the amplitude of the GW becomes smaller than the injected GW signal in about 50\% of their simulations.  This implies that the previous analyses are conservative for GW detection, but it can be regarded as aggressive in terms of upper limits.  
In this way, the putative GW signal and existing constraints can be consistent with each other once we take into account the differences of the priors on the pulsar red noise. 
To claim the detection of the GW signals, however, it is also crucial to establish the quadrupole (Hellings-Downs~\cite{Hellings:1983fr}) nature of the GWs.

Assuming the observed stochastic process is due to the detection of stochastic GW background, the NANOGrav paper~\cite{Arzoumanian:2020vkk} studied the possibility that the GWs are produced from supermassive black hole merger events (e.g., see Ref.~\cite{Sesana:2004sp}).  
Other possibilities for the sources of GWs include cosmic strings~\cite{Ellis:2020ena, Blasi:2020mfx, Buchmuller:2020lbh}, the PBH formation~\cite{Vaskonen:2020lbd, DeLuca:2020agl}, and a phase transition of a dark (hidden) sector~\cite{Nakai:2020oit, Addazi:2020zcj}. 

In this paper, we discuss the possibility that the putative GW signal is the second-order GWs induced by the curvature perturbations that produced solar-mass PBHs.  The main difference from Refs.~\cite{Vaskonen:2020lbd, DeLuca:2020agl} is the mass range of the dominant PBH component.  Ref.~\cite{Vaskonen:2020lbd} concluded that the solar-mass PBHs abundance must be negligible and also that the supermassive black holes may be responsible for the NANOGrav signal.  Ref.~\cite{DeLuca:2020agl} considered a wide spectrum of the curvature perturbations and  studied the possibility that the dark matter abundance is explained by $\mathcal{O}(10^{-14})$ solar mass PBHs and a subdominant abundance of the solar-mass PBHs explain the NANOGrav signal.  Further comparisons with Refs.~\cite{Vaskonen:2020lbd, DeLuca:2020agl} are made in Section~\ref{sec:discussion}. 
We compare the second-order GWs and the NANOGrav result in Section~\ref{sec:GW} and interpret it in terms of PBH parameters in Section~\ref{sec:PBH}. Then, we discuss future tests of the scenario by measuring the stochastic GW background from mergers of solar-mass PBHs in Section~\ref{sec:merger}.  After the discussion in Section~\ref{sec:discussion}, we conclude in Section~\ref{sec:conclusion}.  We adopt the natural unit $\hbar = c = 8\pi G = 1$.

\section{NANOGrav signals and second-order GWs} \label{sec:GW}

NANOGrav measures the strain of the GWs which is assumed to be of the power-law type in the relevant range of the analysis,
\begin{align}
h (f) = A_\text{GWB} \left( \frac{f}{f_\text{yr}} \right)^\alpha, \label{strain}
\end{align}
where $f$ is the frequency, $f_\text{yr} = 3.1 \times 10^{-8} \, \text{Hz}$,  $A_\text{GWB}$ is the amplitude, and $\alpha$ is the slope. 
More directly, they measure the timing-residual cross-power spectral density, whose slope is parametrized as $- \gamma = 2 \alpha - 3$. 
They report preferred ranges of the parameter space spanned by $A_\text{GWB}$ and $\gamma$.

These parameters are related to the energy-density fraction parameter $\Omega_\text{GW}(f) = \rho_\text{GW}(f) / \rho_\text{total}$ in the following way, where $\rho_\text{total}$ is the total energy density of the Universe and the GW energy density is given by $\rho_\text{GW} = \int \text{d} \ln f \, \rho_\text{GW}(f)$:~\cite{Arzoumanian:2018saf}
\begin{align}
\Omega_\text{GW}(f)  = \frac{2 \pi^2 f_\text{yr}^2}{3 H_0^2} A_\text{GWB}^2 \left( \frac{f}{f_\text{yr}} \right)^{5 - \gamma} ,
\end{align}
where $H_0 \equiv 100 h  \, \text{km/s/Mpc}$ is the current Hubble parameter.  

In this paper, we discuss the possibility to explain the putative signal by the secondary, curvature-induced GWs produced at the formation of $\mathcal{O}(1) M_\odot$ PBHs.  For such PBHs, it turns out that $f \gtrsim f_\text{yr}$ does not contribute significantly, and so we consider the frequency range $2.5 \times 10^{-9} \, \text{Hz} \leq f \leq 1.2 \times 10^{-8} \, \text{Hz}$~\cite{Ellis:2020ena, Arzoumanian:2020vkk}, which corresponds to the orange contour of figure 1 of Ref.~\cite{Arzoumanian:2020vkk}.  

The current strength of the second-order, curvature-induced GWs is given by $\Omega_\text{GW}(f) = D \Omega_\text{GW,c}(f)$, where $D = (g_{*}(T)/g_{*,0}) (g_{*,s,0}/g_{*,s}(T))^{4/3} \Omega_\text{r} $ is the dilution factor after the matter-radiation equality time with $\Omega_\text{r}$ being the radiation fraction\footnote{
For simplicity, we assume the Standard Model degrees of freedom and that neutrinos are massless.  $g_{*}(T)$ and $g_{*,s}(T)$ are the effective relativistic degrees of freedom for the energy density and the entropy density, respectively~\cite{Saikawa:2018rcs}.  These are evaluated at the horizon entry of the corresponding mode, while the quantities with the subscript $0$ are evaluated at the present time.
}, and $\Omega_\text{GW,c}(f)$ is the asymptotic value of $\Omega_\text{GW}(f)$ well after the production of the GWs but before the equality time.  This is given by 
\begin{align}
\Omega_\text{GW,c} (f) =& \frac{1}{12}\left( \frac{2 \pi f}{  a H} \right)^2 \int_0^\infty \text{d}t \int_{-1}^{1} \text{d}s \left[ \frac{t (t + 2) (s^2 -1)}{(t+s +1)(t-s+1)}\right]^2 \nonumber \\
& \qquad \qquad \qquad   \times  \overline{I^2 (t,s, k \eta_\text{c})} \mathcal{P}_\zeta \left( \pi (t+s+1)f \right) \mathcal{P}_\zeta \left(\pi (t-s+1)f\right) ,
\end{align}
where $a H$ is the conformal Hubble parameter evaluated at the conformal time $\eta_\text{c}$, $\mathcal{P}_\zeta (k)$ is the dimensionless power spectrum of the primordial curvature perturbations, and $\overline{I^2 (t,s,k\eta_\text{c})}$ is the oscillation average of the kernel function, whose analytic formula has been derived in Refs.~\cite{Espinosa:2018eve, Kohri:2018awv}.  For the recent discussions on gauge (in)dependence, see Refs.~\cite{Gong:2019mui, Tomikawa:2019tvi, DeLuca:2019ufz, Inomata:2019yww, Yuan:2019fwv, Giovannini:2020qta, Lu:2020diy,Chang:2020tji,Ali:2020sfw}. 

For the primordial curvature perturbations, we assume that there is a smooth local peak on top of the quasi-scale-invariant power spectrum measured at the cosmic-microwave-background (CMB) scale.  Such a peak can be approximated by the log-normal power spectrum 
\begin{align}
\mathcal{P}_\zeta (k) = \frac{A_\text{s}}{\sqrt{2 \pi \sigma^2}} \exp \left( - \frac{\left( \ln k/k_* \right)^2}{2 \sigma^2} \right),
\end{align}
 where $k = 2\pi f$ is the wave number, $A_\text{s}$ is the amplitude, $\sigma^2$ is the variance, and $\ln k_{*}$ is the average. (One can match the position of the peak, its height, and its width by the Taylor series expansion. Note that the tail parts do not need to be precisely approximated as the log-normal function.)   
We take $\sigma = 1$ throughout the paper as a simple representative value.  An $\mathcal{O}(1)$ value of $\sigma$ can be expected, e.g., if one assumes that the local feature of $\mathcal{P}_\zeta (k)$ originates from a local feature of the inflaton potential, which can be, e.g., a locally flat part (an approximate inflection point)~\cite{Ezquiaga:2017fvi}, a bump, or a dip~\cite{Mishra:2019pzq} in the single-field case, corresponding to some physical phenomenon occurring in $\mathcal{O}(1)$ e-folding time of the Hubble expansion.\footnote{
There are many models that produce such a locally enhanced peak of $\mathcal{P}_\zeta (k)$.  For constructions in the supergravity or string(-inspired) models, see, e.g., Refs.~\cite{Inomata:2017vxo, Gao:2018pvq, Cicoli:2018asa, Ozsoy:2018flq} and references therein.  Also, the effects of changing $\sigma$ on the second-order GWs and on PBHs are studied, e.g., in Ref.~\cite{Pi:2020otn} and Ref.~\cite{Gow:2020bzo}, respectively. 
}
We treat $A_\text{s}$ and $k_{*}$ as free parameters. These can be translated to the GW parameters $A_\text{GWB}$ and $\gamma$ and to the PBH parameters $f_\text{PBH}$ and $M_\text{PBH}$, which are defined below. 
In the case of the log-normal power spectrum, the full (approximate) analytic formula of $\Omega_\text{GW,c}(f)$ is available~\cite{Pi:2020otn} although we compute it numerically with the aid of extrapolation into the IR tail using the formula of Ref.~\cite{Yuan:2019wwo}. 

\begin{figure}[tbh!]
\centering
\includegraphics{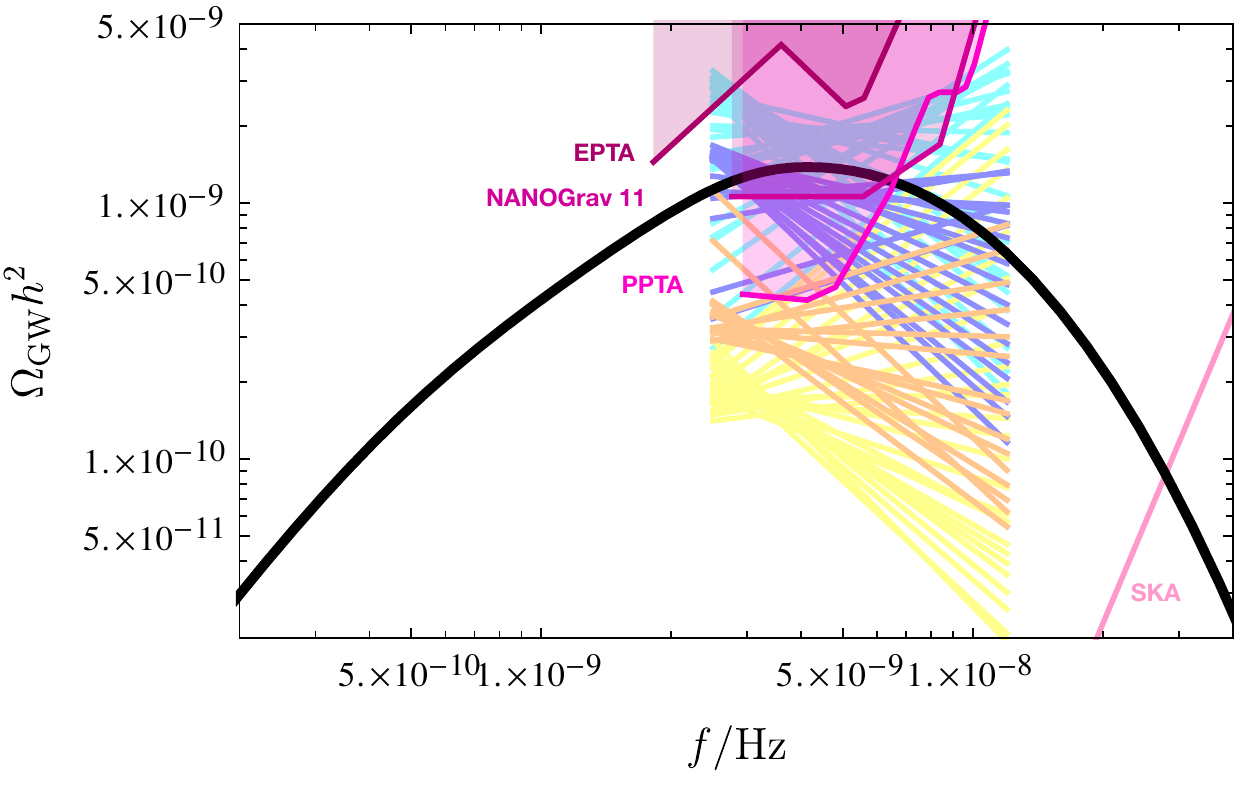}
\caption{\label{fig:omega_GW} Example of the spectrum of the second-order GWs induced by the curvature perturbations that produced PBHs of $M_\text{PBH} = 1 M_\odot$ and $f_\text{PBH} = 1 \times 10^{-4}$ (thick black line).    The power-law lines in the interval $2.5 \times 10^{-9} \, \text{Hz} \leq f \leq 1.2 \times 10^{-8} \, \text{Hz}$ are also shown that correspond to a rough visual guide of the NANOGrav signal range.  The amplitudes and slopes of blue (cyan) and orange (yellow) lines are on the upper and lower $1\sigma$ ($2\sigma$) contours of the NANOGrav signal, respectively.   The previous PTA constraints are shown by shaded regions: EPTA~\cite{Lentati:2015qwp}, NANOGrav 11-yr~\cite{Arzoumanian:2018saf}, and PPTA~\cite{Shannon:2015ect}. The pink line at the bottom right is the prospective constraint of SKA~\cite{5136190}.}
\end{figure}

An example of the spectrum of the second-order GWs is shown as the thick black line in Fig.~\ref{fig:omega_GW}.  Also shown are power-law lines whose amplitude and slope correspond to points on the contours of the NANOGrav favored region on the ($A_\text{GWB}, \gamma$)-plane (the green contours in Fig.~\ref{fig:gamma-A}). The blue and cyan lines correspond to points on the upper half of $1\sigma$ and $2\sigma$ contours, while the orange and yellow lines correspond to points on the lower half  of $1\sigma$ and $2\sigma$ contours, respectively.  The shaded regions are the constraints from the previous PTA observations: EPTA~\cite{Lentati:2015qwp}, NANOGrav 11-yr~\cite{Arzoumanian:2018saf}, and PPTA~\cite{Shannon:2015ect}. 
The pink line at the bottom right is the prospective constraint of SKA~\cite{5136190}.

In the figure, there seems an apparent tension between the NANOGrav 12.5-yr result and the existing PTA constraints.
As mentioned in the introduction, this does not necessarily mean contradiction, but it reflects the intrinsic uncertainties of Bayesian analyses.  
The uniform prior on the red noise for each pulsar (adopted in the existing constraints) tends to pre-assign and overestimate the power in red noise components~\cite{Hazboun:2020kzd}, and the reweighting of the samples of the previous data in accordance with the log-uniform prior indeed weaken the previous constraints~\cite{Arzoumanian:2020vkk, Hazboun:2020kzd}. 
An ongoing joint investigation among the PTA datasets implies a similar tendency to the results of Ref.~\cite{Arzoumanian:2020vkk} also for data other than those of NANOGrav 11-yr~\cite{Arzoumanian:2020vkk} (namely, EPTA and PPTA).  
Therefore, we do not worry too much about the apparent tension between these preexisting PTA constraints and our explanation for the NANOGrav 12.5-yr hint of the GWs in the following analyses.

\section{Implications for the PBH mass and its abundance} \label{sec:PBH}

The relations between the second-order GWs and the properties of PBHs are as follows. 
The GWs are induced by the enhanced curvature perturbations, which also produce PBHs.
The energy density fraction $\beta$ of the PBHs at the formation time, which also has the meaning of the formation probability of a PBH in a given Hubble patch, is calculated in the Press-Schechter formalism~\cite{Press:1973iz}~\footnote{For simplicity, we adopt the Press-Schechter formalism in this paper. However, we would like the readers to refer to Refs.~\cite{Yoo:2018kvb,Suyama:2019npc,Germani:2019zez,Escriva:2020tak,Yoo:2020dkz} for more rigorous treatments.} as
\begin{align}
\beta  =& \int_{\delta_\text{c}}^\infty \text{d} \delta \, \frac{1}{\sqrt{2\pi \sigma_2^2}} \exp \left( - \frac{\delta^2}{2 \sigma_2^2} \right) \simeq \frac{1}{2} \text{Erfc} \left(  \frac{\delta_\text{c}}{\sqrt{2 \sigma_2^2}} \right),
\end{align}
where we have assumed that the primordial curvature perturbations have the Gaussian statistics, $\delta_\text{c}$ is the critical value of the coarse-grained density perturbations that produces a PBH~\cite{Shibata:1999zs, Musco:2004ak, Polnarev:2006aa, Musco:2008hv, Musco:2012au, Nakama:2013ica,Harada:2013epa}, for which we take $\delta_\text{c} = 0.42$~\cite{Harada:2013epa,Harada:2015yda}~\footnote{ \label{fn:delta_c}
For the modified Gaussian window function, it is stated that $\delta_\text{c} = 0.18$ in Table 1 of Ref.~\cite{Young:2019osy} without a detailed derivation.  This may apparently be at odds with a naive expectation that $\delta_\text{c}$ should be higher than in the case of other window functions for the window-function dependence to be suppressed since the modified Gaussian window function enhances the value of $\sigma_2^2$.  For this reason, we take $\delta_\text{c} = 0.42$ as the value used more frequently in the literature. 
}, Erfc is the complementary error function, and the variance $\sigma_2^2$ of the coarse-grained density perturbations is defined as 
\begin{align}
\sigma_2^2 (k) = \frac{16}{81} \int_{-\infty}^\infty \text{d} \ln x \,  w^2 (x)  x^4 \mathcal{P}_\zeta (x k),
\end{align} 
where $w(x)$ is the window function, which we take as the modified Gaussian function $w(x) = \exp( - x^2 /4 )$.  
This window function was introduced in Ref.~\cite{Young:2019osy} and used as one of the two benchmark choices for the window function in Ref.~\cite{Gow:2020bzo}. 
Note that the choice of the window function significantly affects the abundance of the PBHs~\cite{Ando:2018qdb} (see also Ref.~\cite{Tokeshi:2020tjq}) unless compensating parameters for the critical collapse are taken~\cite{Gow:2020bzo}.  We will come back to this point in the discussion section.

The present energy density fraction of PBHs relative to cold dark matter is denoted by $f_\text{PBH} = \rho_\text{PBH}/\rho_\text{CDM}$.  
This is related to $\beta$ as follows, 
\begin{align}
f_\text{PBH} = \int \text{d} \ln M \, \frac{\Omega_\text{m}}{\Omega_\text{CDM}}  \frac{g_{*}(T)}{g_{*}(T_\text{eq})}  \frac{g_{*,s}(T_\text{eq})}{g_{*,s}(T)} \frac{T}{T_\text{eq}}  \epsilon \beta,
\end{align}
where the subscript m and eq denote the non-relativistic matter and the equality time, the temperature $T$ is evaluated at the horizon entry of the corresponding mode $k$, and $\epsilon$ denotes the fraction of the horizon mass that goes into the PBH, which we take $\epsilon = 3^{-3/2}$~\cite{Carr:1975qj}. 
More detailed explanation for PBH formation and parameter dependencies can be found, e.g., in Refs.~\cite{Young:2014ana, Kohri:2018qtx} and in reviews~\cite{Khlopov:2008qy, Carr:2009jm, Carr:2016drx, Sasaki:2018dmp, Carr:2020gox, Carr:2020xqk}. 

We relate $k$ and the horizon mass in the standard way, i.e., using the Friedmann equation.  Note, however, that there is a discrepancy between the average PBH mass $M_\text{PBH}$ and a naive horizon mass corresponding to $k_*$ because of two reasons: the peak position of $\sigma_2^2 (k)$ is smaller than $k_*$, and each PBH mass is $\epsilon$ times smaller than the corresponding horizon mass.  These shifts of peak positions were discussed, e.g., in Ref.~\cite{Wang:2019kaf} and recently emphasized again~\cite{Gow:2020bzo}. 

Concretely, the relation among the wave number $k_*$, the corresponding frequency $f_* = k_* /(2 \pi)$, the corresponding horizon mass $M$, and the average PBH mass $M_\text{PBH}$ is as follows:
\begin{align}
\frac{M_\text{PBH}}{1.0 M_\odot} \simeq  \frac{M}{0.31 M_\odot}  \simeq \left( \frac{k_*}{3.3 \times 10^6 \, \text{Mpc}^{-1}} \right)^{-2} \simeq \left( \frac{f_*}{5.0 \times 10^{-9} \, \text{Hz}} \right)^{-2} .  \label{linear_relation}
\end{align}

\begin{figure}[tbh!]
\centering
\includegraphics{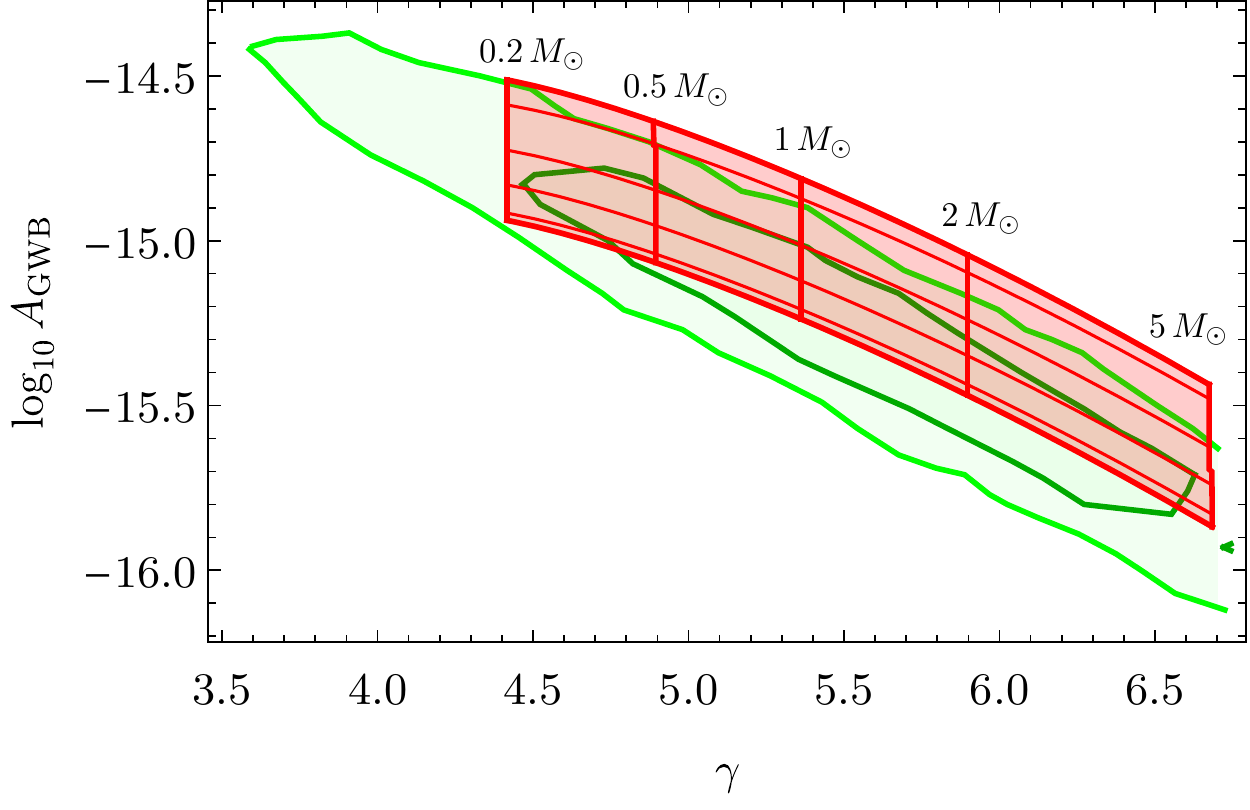}
\caption{\label{fig:gamma-A} Parameter scan in the range $0.015 \leq A_\text{s} \leq 0.040$ and $0.2 \leq M_\text{PBH}/M_\odot \leq 5$ shown as the red shaded region.  A larger $A_\text{s}$ corresponds to a larger  $A_\text{GWB}$, and a larger $M_\text{PBH}$ corresponds to a larger $\gamma$. The thin red lines correspond to $f_\text{PBH} = 10^{-1}$, $10^{-4}$, $10^{-7}$, and $10^{-10}$ from top to bottom.  The $1\sigma$ and $2\sigma$ NANOGrav contours are also shown. }
\end{figure}

We vary the scalar amplitude in the range $0.015 \leq A_\text{s} \leq 0.040$ and the average PBH mass in the range $0.2 \leq M_\text{PBH} / M_\odot \leq 5$.  The resultant $\Omega_\text{GW} h^2$ is fitted by a power-law line in the aforementioned range $2.5 \times 10^{-9} \, \text{Hz} \leq f \leq 1.2 \times 10^{-8} \, \text{Hz}$ to extract the amplitude of the GW strain $A_\text{GWB}$ and the slope $\gamma$. Note that $A_\text{GWB} \propto A_\text{s}$, but it also depends on $k_\text{*}$ (or $M_\text{PBH}$) since the pivot scale is fixed to $f_\text{yr}$ (see eq.~\eqref{strain}).  The result is shown in Fig.~\ref{fig:gamma-A}. From the figure, we see that a large fraction of the scanned parameter space can explain the NANOGrav signal.

\begin{figure}[tbh!]
\centering
\includegraphics{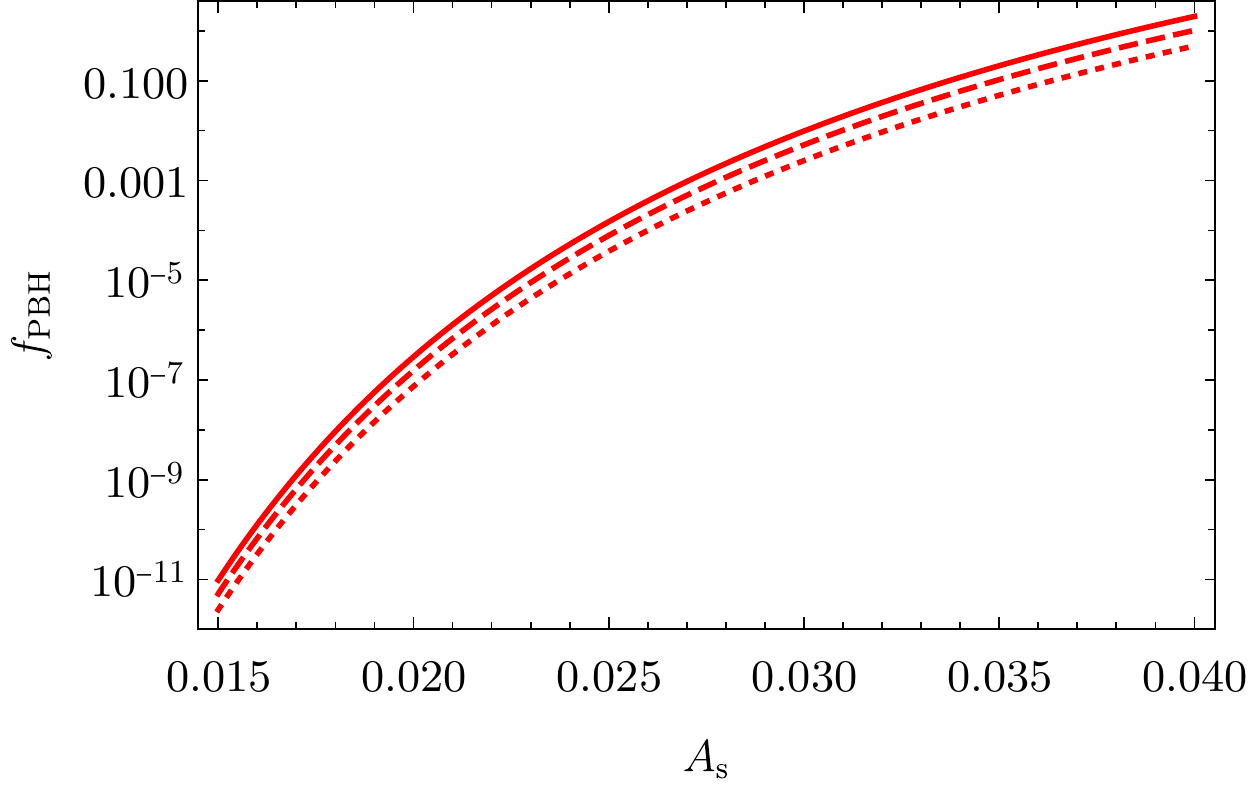}
\caption{\label{fig:A-f} Relation between the scalar amplitude $A_\text{s}$ and the PBH abundance $f_\text{PBH}$ for $M_\text{PBH}/M_\odot = 0.2$ (top, solid), $1$ (middle, dashed), and $5$ (bottom, dotted).} 
\end{figure}

The scanned parameter range for $A_\text{s}$ corresponds to that of the PBH abundance $f_\text{PBH}$ as shown in Fig.~\ref{fig:A-f}.  The upper and lower ends correspond to $M_\text{PBH} = 0.2 M_\odot$ and $5 M_\odot$, respectively. 

\begin{figure}[tbh!]
\centering
\includegraphics{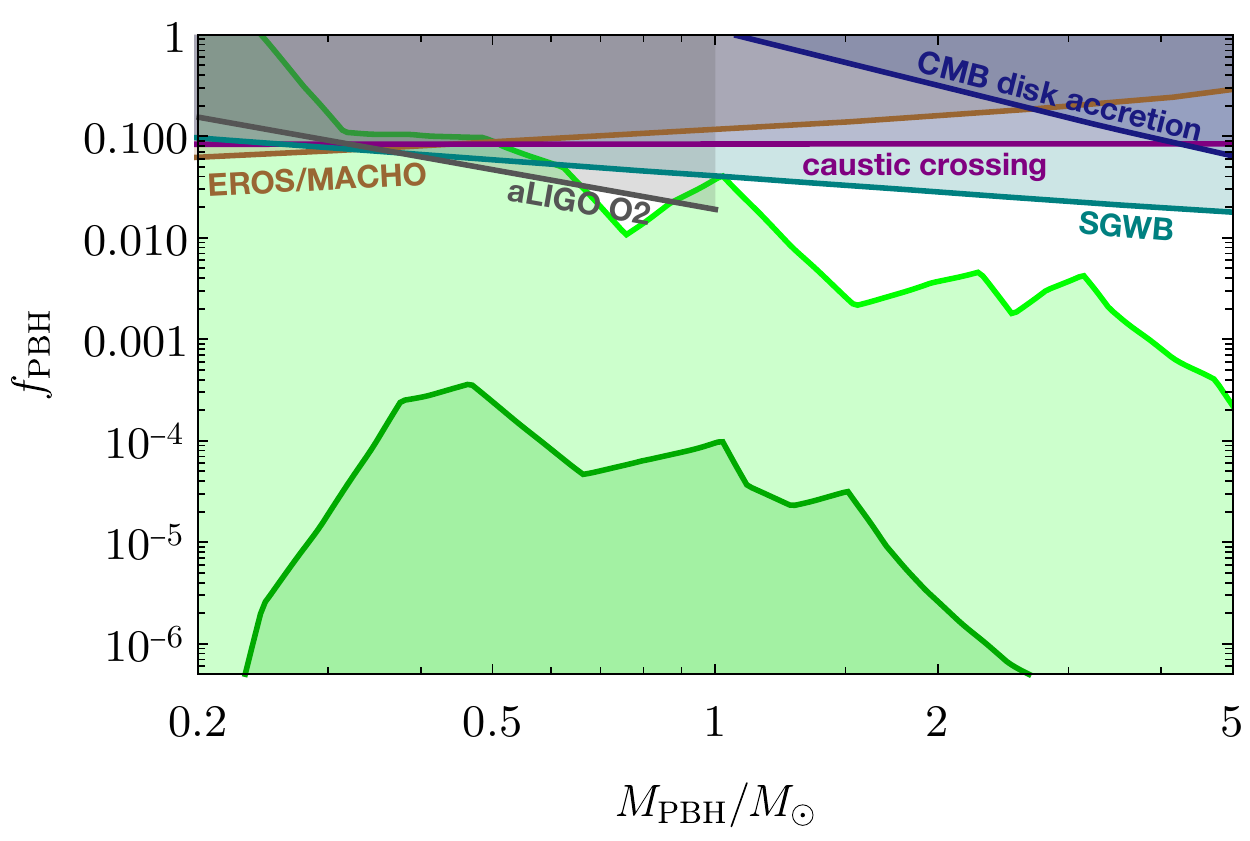}
\caption{\label{fig:M-f} NANOGrav contours (green) on the plane of the average PBH mass $M_\text{PBH}$ and the PBH abundance $f_\text{PBH}$. The dark shaded regions at the top are constraints from EROS-2~\cite{Tisserand:2006zx} and MACHO~\cite{Allsman:2000kg} (brown), caustic crossing~\cite{Oguri:2017ock} (purple), Advanced LIGO O2 (subsolar mass range)~\cite{Authors:2019qbw} (gray), Advanced LIGO non-detection of the stochastic GW background~\cite{Wang:2016ana, Chen:2019irf} (cyan), and the $E$-mode polarization of the CMB due to the disk-shaped gas accretion~\cite{Poulin:2017bwe} (blue). } 
\end{figure}

Combining the information in Figs.~\ref{fig:gamma-A} and \ref{fig:A-f}, one can map the NANOGrav contours onto the PBH parameter space ($M_\text{PBH}$, $f_\text{PBH}$), which are shown as the green contours in Fig.~\ref{fig:M-f}.  The non-smoothness of the contours largely originates from the non-smoothness of the original NANOGrav contours.  The uncertainty of extracting the data from the original contours is magnified in this figure compared to Fig.~\ref{fig:gamma-A}.  Therefore, the $1\sigma$ and $2\sigma$ boundary has an uncertainty of very roughly an order of magnitude. 

Fig.~\ref{fig:M-f} shows that the PBH mass should be around a solar mass to explain the NANOGrav signal. Also,  it shows that $f_\text{PBH}$ close to unity is disfavored, but $f_\text{PBH} \sim 0.1$ is within the $2\sigma$ contour depending on the value of $M_\text{PBH}$. 

  A part  of such  regions is excluded by existing constraints shown by shaded regions at the top of the figure.  These include the microlensing constraints by EROS/MACHO collaborations~\cite{Tisserand:2006zx, Allsman:2000kg}, the caustic crossing constraint~\cite{Oguri:2017ock}, Advanced LIGO constraints on the subsolar mass range (individual events~\cite{Authors:2019qbw} and superposition of events~\cite{Wang:2016ana, Chen:2019irf}), and the constraints due to photo-emission during gas accretion onto PBHs~\cite{Ali-Haimoud:2016mbv, Poulin:2017bwe, Serpico:2020ehh}.  There are many subdominant but independent and complementary constraints around this mass range (see Ref.~\cite{Carr:2020gox}).
There is also the LIGO/Virgo constraints on supersolar mass range~\cite{Ali-Haimoud:2017rtz, Vaskonen:2019jpv}.  Ref.~\cite{Vaskonen:2019jpv} implies a substantial dependence on the width of the mass function, so we do not include it in Fig.~\ref{fig:M-f}.

\section{Testing the scenario with the GWs from mergers} \label{sec:merger}

The solar-mass PBH possibility for NANOGrav can be tested by the detection of stochastic GW background from the superposition of binary solar-mass PBH merger events. The GW spectrum is obtained as 
\begin{align}
\Omega_\text{GW}^{\text{merger}} (f) = \frac{f}{3 H_0^2} \int_0^{\frac{f_\text{cut}}{f}-1} \text{d} z \, \frac{R (z) }{(1+z) H(z)} \frac{\text{d}E_\text{GW}}{\text{d} f_\text{s}} ,   \label{Omega_GW_merger}
\end{align}
where $f_\text{cut}$ ($=\mathcal{O} (1/M_\text{PBH})$) is the UV cutoff frequency at the source frame (i.e., without the redshift factor) (see Refs.~\cite{Cai:2019cdl, Inomata:2020lmk} for the IR ``cutoff'' frequency), $f_\text{s}$ is the frequency at the source frame, $z$ is the redshift, $R$ is the comoving merger rate, and $E_\text{GW}$ is the energy of the GWs at the source frame. The expressions of $f_\text{cut}$, $R$, and $\text{d}E_\text{GW}/\text{d} f_\text{s}$ are found in Appendices B and C of Ref.~\cite{Wang:2019kaf}. See also Refs.~\cite{Wang:2016ana, Sasaki:2016jop, Sasaki:2018dmp, Ajith:2007kx, Ajith:2009bn} for more details.  The frequency $f_\text{cut}$ is just the maximal cutoff appearing around the end of the merger process.

\begin{figure}[tbh!]
\centering
\includegraphics{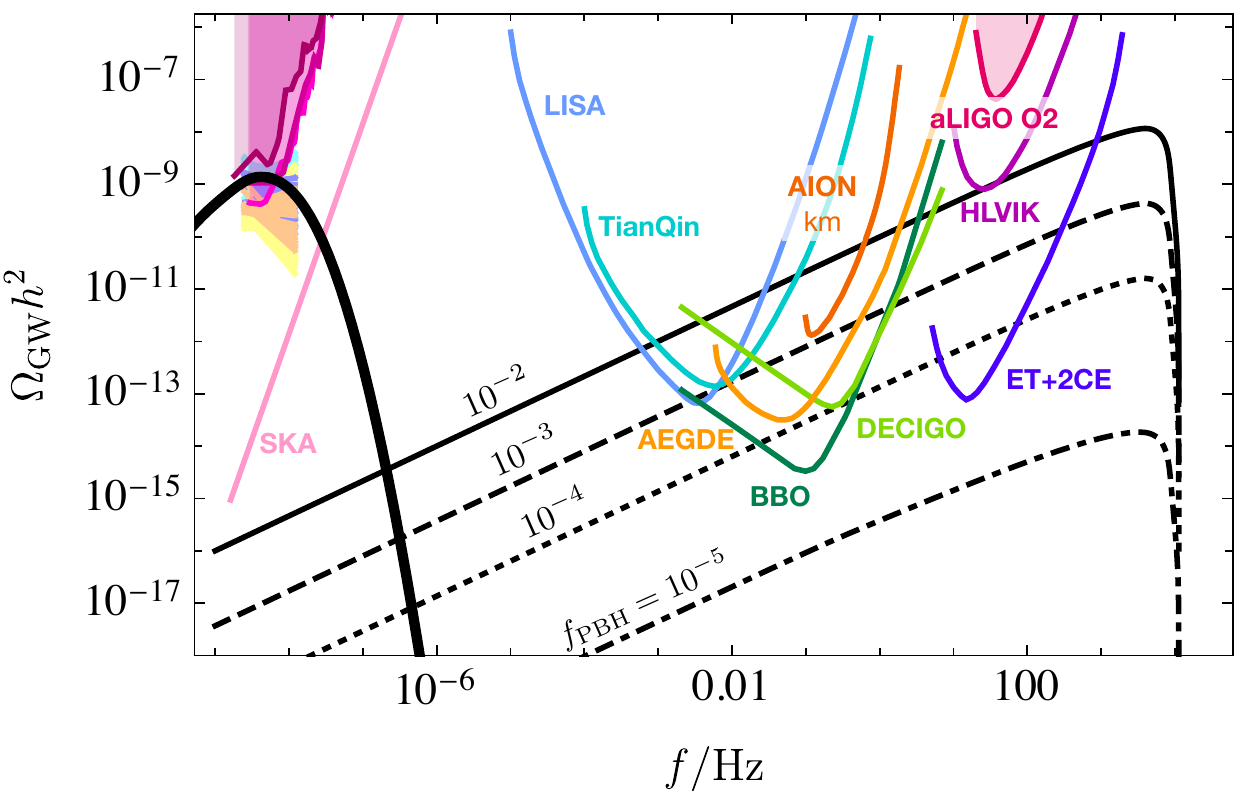}
\caption{\label{fig:merger} GW spectrum from the superposition of binary PBH merger events (thin black) with $M_\text{PBH} = 1 M_\odot$ and $f_\text{PBH} = 10^{-2}$ (solid), $10^{-3}$ (dashed),  $10^{-4}$ (dotted), and $10^{-5}$ (dot-dashed). Future prospects of various GW observations are also shown: SKA~\cite{5136190}, LISA~\cite{2017arXiv170200786A}, TianQin~\cite{Luo:2015ght, Mei:2020lrl}, BBO~\cite{Harry:2006fi}, DECIGO~\cite{Seto:2001qf}, AION~\cite{Badurina:2019hst}, AEDGE~\cite{Bertoldi:2019tck},  Advanced LIGO Hanford and Livingston~\cite{TheLIGOScientific:2014jea} combined with Advanced Virgo~\cite{TheVirgo:2014hva} as well as LIGO India~\cite{LIGOIndia, Unnikrishnan:2013qwa} and KAGRA~\cite{Somiya:2011np, Aso:2013eba} (HLVIK), and Einstein Telescope~\cite{Punturo:2010zz} and two third-generation Cosmic Explorers~\cite{Evans:2016mbw} (ET+2CE).
The shaded red region is the Advanced LIGO O2 constraint~\cite{LIGOScientific:2019vic}.
Sensitivity curves have been read from Refs.~\cite{Moore:2014lga, Badurina:2019hst, Ellis:2020ena, Perigois:2020ymr}.
    The top side of the figure is the upper bound $\Omega_\text{GW}h^2 < 1.8 \times 10^{-6}$ from the (non-)adiabatic $N_\text{eff}$ bound of big-bang nucleosynthesis~\cite{Kohri:2018awv}. The existing PTA constraints and NANOGrav power-law guides are also shown as in Fig.~\ref{fig:omega_GW}.}
\end{figure}

The result is shown in Fig.~\ref{fig:merger} as the black lines where $M_\text{PBH} = 1 M\odot$ and $f_\text{PBH}= 10^{-2}$ (solid), $10^{-3}$ (dashed),  $10^{-4}$ (dotted), and $10^{-5}$ (dot-dashed).  Various prospective constraints (see the caption)\footnote{
Though not shown in the figure, see also the following references for related experiments: ALIA~\cite{Bender:2013nsa}, ELGAR~\cite{Canuel:2019abg}, MAGIS~\cite{Graham:2017pmn, Coleman:2018ozp}, MIGA~\cite{Canuel:2017rrp}, Taiji~\cite{Guo:2018npi}, and ZAIGA~\cite{Zhan:2019quq}.  
} as well as the lines in Fig.~\ref{fig:omega_GW} are also shown.   
We do not show the $M_\text{PBH}$ dependence in the figure, but the spectra shift to the left as $M_\text{PBH}$ increases.  Eq.~\eqref{linear_relation} clearly shows that the characteristic frequency $f_*$ of the second-order GWs scales as $M_\text{PBH}^{-1/2}$, whereas the counterpart for the GWs from mergers scales as $f_\text{cut} \sim M_\text{PBH}^{-1}$ (see the text below eq.~\eqref{Omega_GW_merger}) as demonstrated in Ref.~\cite{Wang:2019kaf}. 
Note that the thick black line corresponds to the second-order GWs for $M_\text{PBH} = 1 M_\odot$ and $f_\text{PBH} = 10^{-4}$, but the $f_\text{PBH}$ dependence is weak (see Fig.~\ref{fig:A-f}).
The top end of the figure is the upper bound  $\Omega_\text{GW}h^2 < 1.8 \times 10^{-6}$~\cite{Kohri:2018awv} from the fact that the GWs contribute to the effective number of neutrinos $N_\text{eff}$ and affect the big-bang nucleosynthesis. 
We can see from the figure that a large part of the parameter space can be probed by the future GW observations.

\section{Discussion} \label{sec:discussion}

Our results depend on various assumptions.  Some of them have been
already stated, but we emphasize them again.  First, we do not
consider the effect of the critical collapse~\cite{Choptuik:1992jv,
  Niemeyer:1997mt, Green:1999xm, Carr:2016drx} since it occurs only
when the spherical symmetry is precisely respected.  It is clear that
the rare high-peak has approximately the spherical
shape~\cite{Bardeen:1985tr}, but the spherical symmetry must be
realized to high precision for the critical collapse to
happen~\cite{Shibata:1999zs}.
On the other hand, Refs.~\cite{Vaskonen:2020lbd, DeLuca:2020agl} include the effect of the critical collapse. 
It will be interesting to compare our results with an analysis including the critical collapse effect using a consistent parameter set~\cite{Gow:2020bzo}. 
In our preliminary study, we found a qualitatively similar feature that $f_\text{PBH}$ tends to become larger than those reported in Refs.~\cite{Vaskonen:2020lbd, DeLuca:2020agl}.

Second, we have chosen the modified Gaussian window function, whose width is twice as large as the standard Gaussian window function.  This boosts the value of $f_\text{PBH}$ for a given value of $A_\text{s}$.  This may be the largest difference compared to Refs.~\cite{Vaskonen:2020lbd, DeLuca:2020agl} in which much smaller $f_\text{PBH}$'s were reported.  

Third, we have not taken into account the nonlinear relation between
the primordial curvature perturbations and the density perturbations
(see Refs.~\cite{Yoo:2018kvb, Kawasaki:2019mbl}). This inevitably leads
to non-Gaussianity of the density
perturbations~\cite{Kawasaki:2019mbl}.  Also, the inclusion of the
intrinsic non-Gaussianity of the primordial curvature perturbations
significantly affects $f_\text{PBH}$~\cite{Byrnes:2012yx, Young:2013oia, Unal:2020mts}.
It also affects the second-order GWs~\cite{Cai:2018dig, Yuan:2020iwf, Garcia-Bellido:2017aan, Unal:2018yaa, Nakama:2016gzw, Unal:2020mts}.

Fourth, we have not included the transfer function of the curvature perturbations in the definition of $\sigma_2^2$.  This is preferred in Ref.~\cite{Young:2019osy}.  
  If we include the transfer function, however, $\sigma_2^2$ will reduce by ``several'' percent.  This reduces $f_\text{PBH}$ non-negligibly. 

It is also worth mentioning that we have not taken into account the softening of the equation-of-state during the phase transition/crossover of quantum chromodynamics (QCD).  See Refs.~\cite{DeLuca:2020agl, Byrnes:2018clq, Jedamzik:1996mr} for its enhancement effect on the PBH abundance $f_\text{PBH}$ for a given scalar amplitude $A_\text{s}$.  Depending on the boost factor, this may realize a better fit for the NANOGrav signal simultaneously with stronger and more easily detectable  GWs from mergers of the solar-mass binary PBHs. The softening also slightly affects the spectrum of the second-order GWs~\cite{Hajkarim:2019nbx}.    

We discussed a possible detection of the PBHs with the masses
of ${\cal O}(1) M_{\odot}$ only by a future interferometer-type GW
observations in Section~\ref{sec:merger}. Complementarily, however,
we can also measure such PBHs by the future optical/IR telescopes
through microlensing events, e.g., Subaru HSC towards M31 for 10
year observations~\cite{Takada:2019} or by the future precise CMB
observations of $E$- and $B$-mode polarization due to photon
emission from an accretion disk around a PBH, e.g., by
LiteBIRD~\cite{Hazumi:2019lys} or CMB-S4~\cite{Abazajian:2019eic}.

\section{Conclusion} \label{sec:conclusion}
In this paper, we have interpreted the recently reported NANOGrav 12.5-yr excess of the timing-residual cross-power spectral density in the low-frequency part as a stochastic GW background.  We conclude that, under our assumptions, the second-order GWs induced by the curvature perturbations that produced a substantial amount of $\mathcal{O}(1)$ solar-mass PBHs can explain the NANOGrav stochastic GW signal.  In particular, the abundance of the PBHs can be sufficiently large so that future GW observations can test this possibility by measuring the stochastic GW background produced by mergers of the solar-mass PBHs.  
This is nontrivial since the suitable scalar amplitude $A_\text{s}$ could a priori produce too many PBHs that are excluded by existing observational constraints or too few PBHs that do not lead to the detectable stochastic GW background from merger events.  Similarly, for a given $f_\text{PBH}$, the second-order GWs could be too strong or weak.  Since the relation between $A_\text{s}$ and $f_\text{PBH}$ depends crucially on the ambiguity for the choice of the windows function as discussed in the previous section, a further study to refine the PBH formation criterion is necessary.

\section*{Note Added}
Taking into account the uncertainties of PBH abundance calculations, i.e., the different choices of the window function, the value of $\delta_\text{c}$ (see footnote~\ref{fn:delta_c}), etc., our results are largely consistent with those of Ref.~\cite{DeLuca:2020agl}~\cite{Franciolini}. 
The difference from Ref.~\cite{Vaskonen:2020lbd} is also discussed in the note added in Ref.~\cite{Vaskonen:2020lbd}.  In our paper, we do not claim that $\mathcal{O}(30) M_\odot$ PBHs responsible for the LIGO/Virgo events can explain the NANOGrav signal.

\section*{Acknowledgments}
K.K.~thanks Misao Sasaki and Shi Pi for useful discussions. 
We thank Valerio De Luca, Gabriele Franciolini, and Antonio Riotto for useful discussions.
This work was supported by JSPS KAKENHI
Grant Number JP17H01131 (K.K.), MEXT KAKENHI Grant Numbers JP19H05114 (K.K.) and JP20H04750 (K.K.), and IBS under the project code, IBS-R018-D1 (T.T.).

\small

\bibliographystyle{utphys}
\bibliography{NANOGrav.bib}

\end{document}